\documentclass[twocolumn,prb,9pt,superscriptaddress,amsfonts]{revtex4-1}

\usepackage[pdftex]{graphicx}
\usepackage{xcolor}
\usepackage[pdftex,bookmarks=true,bookmarksopen,bookmarksnumbered,
colorlinks,
linkcolor=blue,
citecolor=blue]{hyperref}
\usepackage{bm}
\usepackage{mathtools}
\usepackage{graphicx}
\usepackage{graphics}
\usepackage{dcolumn}
\usepackage{float}
\usepackage{epsfig}
\usepackage{epstopdf} 
\usepackage{color}
\usepackage{amsmath}
\usepackage{textcomp}
\usepackage{multirow}
\usepackage{array}
\usepackage{newfloat}

\DeclareFloatingEnvironment[
fileext=los,
name=Figure,
placement=tbhp,
within=none,
]{suppfig}

\renewcommand{\figurename}{Figure}
\makeatletter
\renewcommand*{\fnum@figure}{{\normalfont\bfseries \figurename~\thefigure}}
\renewcommand*{\@caption@fignum@sep}{ $|$}
\newcommand*{\rom}[1]{\expandafter\@slowromancap\romannumeral #1@}
\makeatother

\pdfpageattr {/Group << /S /Transparency /I true /CS /DeviceRGB>>} 

\begin{document}
	
	\title{Coherent spin control of s-, p-, d- and f-electrons in a silicon quantum dot}
	
	\author{R. C. C. Leon}
	\email[r.leon@student.unsw.edu.au]{}
	\affiliation{Centre for Quantum Computation and Communication Technology, School of Electrical Engineering and Telecommunications, The University of New South Wales, Sydney, NSW 2052, Australia.}
	\author{C. H. Yang}
	\affiliation{Centre for Quantum Computation and Communication Technology, School of Electrical Engineering and Telecommunications, The University of New South Wales, Sydney, NSW 2052, Australia.}
	\author{J. C. C. Hwang}	
	\altaffiliation{Current address: Research and Prototype Foundry, The University of Sydney, Sydney, NSW 2006, Australia.}
	\affiliation{Centre for Quantum Computation and Communication Technology, School of Electrical Engineering and Telecommunications, The University of New South Wales, Sydney, NSW 2052, Australia.}
	\author{J. Camirand Lemyre}
	\affiliation{Institut Quantique et D\'epartement de Physique, Universit\'e de Sherbrooke, Sherbrooke, Qu\'ebec J1K 2R1, Canada}
	\author{T. Tanttu}
	\affiliation{Centre for Quantum Computation and Communication Technology, School of Electrical Engineering and Telecommunications, The University of New South Wales, Sydney, NSW 2052, Australia.}
	\author{W. Huang}
	\affiliation{Centre for Quantum Computation and Communication Technology, School of Electrical Engineering and Telecommunications, The University of New South Wales, Sydney, NSW 2052, Australia.}
	\author{K. W. Chan}
	\affiliation{Centre for Quantum Computation and Communication Technology, School of Electrical Engineering and Telecommunications, The University of New South Wales, Sydney, NSW 2052, Australia.}
	\author{K. Y. Tan}
	\affiliation{QCD Labs, COMP Centre of Excellence, Department of Applied Physics, Aalto University, 00076 Aalto, Finland}
	\author{F. E. Hudson}
	\affiliation{Centre for Quantum Computation and Communication Technology, School of Electrical Engineering and Telecommunications, The University of New South Wales, Sydney, NSW 2052, Australia.}
	\author{K. M. Itoh}
	\affiliation{School of Fundamental Science and Technology, Keio University, 3-14-1 Hiyoshi, Kohokuku, Yokohama 223-8522, Japan.}
	\author{A. Morello}
	\affiliation{Centre for Quantum Computation and Communication Technology, School of Electrical Engineering and Telecommunications, The University of New South Wales, Sydney, NSW 2052, Australia.}
	\author{A. Laucht}
	\affiliation{Centre for Quantum Computation and Communication Technology, School of Electrical Engineering and Telecommunications, The University of New South Wales, Sydney, NSW 2052, Australia.}
	\author{M. Pioro-Ladri\`ere}
	\affiliation{Institut Quantique et D\'epartement de Physique, Universit\'e de Sherbrooke, Sherbrooke, Qu\'ebec J1K 2R1, Canada}
	\affiliation{Quantum Information Science Program, Canadian Institute for Advanced Research, Toronto, ON, M5G 1Z8, Canada}
	\author{A. Saraiva}
	\email[a.saraiva@unsw.edu.au]{}
	\affiliation{Centre for Quantum Computation and Communication Technology, School of Electrical Engineering and Telecommunications, The University of New South Wales, Sydney, NSW 2052, Australia.}
	\author{A. S. Dzurak}
	\email[a.dzurak@unsw.edu.au]{}
	\affiliation{Centre for Quantum Computation and Communication Technology, School of Electrical Engineering and Telecommunications, The University of New South Wales, Sydney, NSW 2052, Australia.}
	
	
	\begin{abstract}
		\textbf{Once the periodic properties of elements were unveiled, chemical bonds could be understood in terms of the valence of atoms. Ideally, this rationale would extend to quantum dots, often termed artificial atoms, and quantum computation could be performed by merely controlling the outer-shell electrons of dot-based qubits. Imperfections in the semiconductor material, including at the atomic scale, disrupt this analogy between atoms and quantum dots, so that real devices seldom display such a systematic many-electron arrangement. We demonstrate here an electrostatically-defined quantum dot that is robust to disorder, revealing a well defined shell structure. We observe four shells (31 electrons) with multiplicities given by spin and valley degrees of freedom. We explore various fillings consisting of a single valence electron -- namely 1, 5, 13 and 25 electrons -- as potential qubits, and we identify fillings that yield a total spin-1 on the dot. An integrated micromagnet allows us to perform electrically-driven spin resonance (EDSR). Higher shell states are shown to be more susceptible to the driving field, leading to faster Rabi rotations of the qubit. We investigate the impact of orbital excitations of the p- and d-shell electrons on single qubits as a function of the dot deformation. This allows us to tune the dot excitation spectrum and exploit it for faster qubit control. Furthermore, hotspots arising from this tunable energy level structure provide a pathway towards fast spin initialisation. The observation of spin-1 states may be exploited in the future to study symmetry-protected topological states in antiferromagnetic spin chains and their application to quantum computing.
		}
	\end{abstract}
	
	\maketitle
	
	Qubit architectures based on electron spins in gate-defined silicon quantum dots benefit from a high level of controllability, where single and multi-qubit coherent operations are realised solely with electrical and magnetic manipulation. Furthermore, their direct compatibility with silicon microelectronics fabrication offers unique scale-up opportunities\cite{Vandersypen2017}. However, fabrication reproducibility and disorder pose challenges for single electron quantum dots. Even when the single-electron regime is achievable, the last electron often is confined in a very small region, limiting the effectiveness of electrical control and interdot tunnel coupling. Many-electron quantum dots were proposed as a qubit platform decades ago\cite{Hu2001}, with the potential of resilience to charge noise\cite{barnes2011screening,bakker2015validity} and a more tunable tunnel coupling strength to other qubits\cite{Harvey-Collard2017}. 
	In the multielectron regime, the operation of a quantum dot qubit is more sensitive to its shape. If it is axially symmetric, the orbital energy levels will be quasi-degenerate\cite{tarucha1996shell,kouwenhoven1997excitation,rontani2006full}, which is detrimental for quantum computing. On the contrary, if the quantum dot is very elongated, a regular shell structure will not form, and the valence electron will not operate as a simple spin-$1/2$ system\cite{Hu2001,Deng2018}.

	\begin{figure*}
		\centering
		\includegraphics[width=\linewidth]{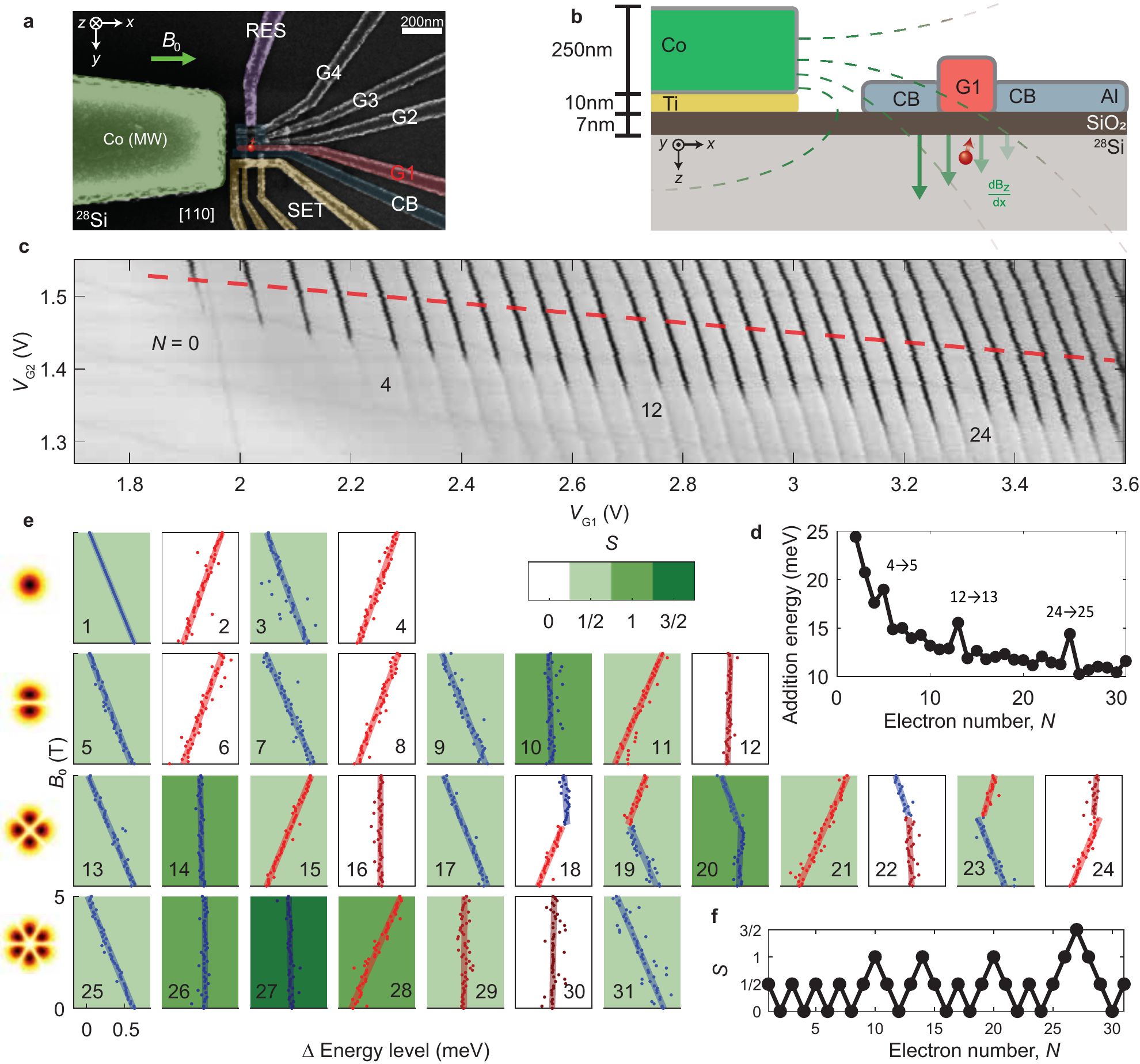}
		\caption{\textbf{Device overview and electron occupancy measurement.} 
			\textbf{a}, False-coloured SEM image of a nominally identical device to that reported here. A quantum dot is formed under gate G1 (red), in the location marked by the red symbol. Gate RES is connected to an n-doped reservoir to load/unload electrons to/from the quantum dot, with tunnel rates controlled by G2, G3 and G4. Gate CB serves as a confinement barrier. The cobalt (Co) structure at the left of the image acts as both a micromagnet and electrode for EDSR control (green).
			\textbf{b}, Cross-sectional schematic of the device, fabricated on a purified Silicon-28 epi-layer (800 ppm).
			\textbf{c}, Charge stability map of the quantum dot at $B_0=0$T, produced by plotting the pulsed lock-in signal from SET sensor $I_\textrm{pulse}$ vs $V_\textrm{G1}$ and $V_\textrm{G2}$. A square wave with peak-to-peak amplitude of 2 mV is applied to G1 for lock-in excitation. Dynamic compensation is applied to the SET sensor to maintain a high readout sensitivity. Electron numbers $N$ for full shells are marked on the diagram.
			\textbf{d}, Charging energies along the red line in \textbf{(c)} in the tightly confined regime.
			\textbf{e}, Magnetospectroscopy of the first 31 electrons occupied in the quantum dot, up to $B_0=5$ T, with background colour of each plot representing spin state $S$ at $B_0=0$ T. Change in addition energies with magnetic field are measured and fitted with straight lines. Since the charging energy is measured only from the second electron, the first electron is depicted by a straight line with no data. Each row of the array of plots belongs to the same shell, while each column has the same number of valence electrons in its outershell. The cartoon on the left gives an example of the electron wavefunction for each shell.
			\textbf{f}, Spin state of each electron occupancy extracted from \textbf{(e)}.
		}\label{SEMs}
	\end{figure*}
	
	\bigskip
	
	\section*{Filling s-, p-, d- and f-orbitals in a silicon quantum dot} 
	The scanning electron microscope (SEM) image in Fig.~\ref{SEMs}a shows a silicon semiconductor-metal-oxide (Si-MOS) device that forms a quantum dot at the Si/SiO$_2$ interface under gate G1, separated from the reservoir by a barrier that is controlled by gate G2 - see Fig.~\ref{SEMs}b for a cross-sectional representation. We first study the electronic structure of the dot from its charge stability diagram, using the technique from Ref.~\onlinecite{Yang2011}, which maps out each electron transition between quantum dot and reservoir as a function of gate potentials. Figure~\ref{SEMs}c shows an extremely regular set of electron transitions, revealing a quantum dot that can be occupied by up to 31 electrons with no significant evidence of disorder related to random fixed charge states in the SiO$_2$ or at the Si/SiO$_2$ interface. This occupancy range is slightly better than other devices based on similar technology~\cite{Yang2013}. Additional charge transitions in Fig.~\ref{SEMs}c (faint nearly-horizontal lines) arise from states between the reservoir and the quantum dot and do not affect the qubit operation. Lowering the voltage of gate G2 confines the quantum dot further and changes its eccentricity in the x-y plane. 
	
	Following the red dashed line in Fig.~\ref{SEMs}c allows us to investigate the addition energies, \textit{i.e.}, the energy necessary to add the $N$-th electron to a dot that contains $N$-1 electrons, as plotted in Fig.~\ref{SEMs}d. The first noticeable effect is that the charging energy is roughly inversely proportional to the number of electrons, which is a consequence of the dot size becoming larger as the dot fills up. Furthermore, very distinct peaks appear at transitions $4\rightarrow 5$, $12\rightarrow 13$, and $24\rightarrow 25$. To understand the significance of these electron numbers, one may refer to the Fock-Darwin energy levels \cite{fock1928bemerkung,darwin1931diamagnetism}, where the internal spin ($\uparrow,\downarrow$) and valley ($v_+, v_-$) quantum numbers give the multiplicity of each orbital state in a two-dimensional quantum dot. As a result, a full shell is formed when there are 4, 12 and 24 electrons in the 2D quantum dot, and so an extra energy, corresponding to the orbital level splitting, must be supplied in order to begin filling the next shell. The filling of three complete electron shells has previously been observed in a GaAs quantum dot\cite{tarucha1996shell}, where the single-valley nature of the semiconductor leads to a filled third shell at $N$=12 electrons, but until now has not been observed in a silicon device. The observed shell filling is analogous to the aufbau principle of atomic physics, that allows us to construct the electronic structure of many-electron atoms in terms of occupation of the atomic electron levels from bottom up. 
	
	As well as the large jumps in addition energy observed after complete shells are filled, a finer structure at intermediate fillings is also present due to the valley splitting $\Delta_\textrm{VS}$ \cite{Zwanenburg2013}, the energy difference between excitations along the major and minor axes of the elliptical quantum dot \cite{Ngo2006} $\Delta_{xy}$, and electronic quantum correlations\cite{harting2000interplay}, dominated by the exchange coupling $J$. These energy scales are much smaller than the shell excitation, so that we can identify each set of levels by a principal quantum number. Each shell is spanned by the valley\cite{lim2011spin,borselli2011measurement,Yang2012}, spin and azimuthal\cite{Jacak1998} quantum numbers. For this particular quantum dot, $\Delta_\textrm{VS}$ and $\Delta_{xy}$ may be estimated~\cite{borselli2011measurement,Yang2013} and both are of the order of hundreds of $\mu$eV, which is consistent with typical observations for quantum dots with similar designs~\cite{Yang2012}. Since both splittings are similar in magnitude, it is difficult to label the inner shell structure based solely on the addition energy diagram.
	
	Magnetospectroscopy of the electron transitions (Fig.~\ref{SEMs}e) reveals the spin dependency as a function of external magnetic field strength $B_0$ for each electron occupancy, with cumulative spin state $S$ presented in Fig. \ref{SEMs}f. At lower electron occupancies, $S$ alternates between $0$ and $\frac{1}{2}$, indicating that the sequential electron loading favours anti-parallel spin states, implying $J\ll\Delta_{xy},\Delta_\textrm{VS}$. As the electron numbers increases, Hund's rule applies as some of the electrons are loaded as parallel spin ($S=1$ or $\frac{3}{2}$ states), indicating $J>\Delta_{xy},\Delta_\textrm{VS}$ in these cases.
	
	The observation of S=1 spin states is potentially significant, in the context of the study of symmetry-protected topological phases of $S=1$ spin chains with antiferromagnetic Heisenberg coupling. As conjectured by Haldane \cite{Haldane1983}, such $S=1$ spin chains possess a fourfold degenerate ground state, protected by a topological gap to higher excited states. Finite-length chains exhibit fractionalized $S=1/2$ states at their ends, which could be exploited for robust quantum computing schemes \cite{Brennen2008,Bartlett2010,Miyake2010}. The experimental realization of controllable $S=1$ Haldane chains, however, has remained a formidable challenge \cite{Senko2015}. In semiconductor quantum dots, methods to locally control and read out chains of spins are now mature. Engineering $S=1$ with the natural Heisenberg exchange interaction in this system might open exciting opportunities for future studies in this field. 
	
	\begin{figure*}
		\centering
		\includegraphics[width=\linewidth]{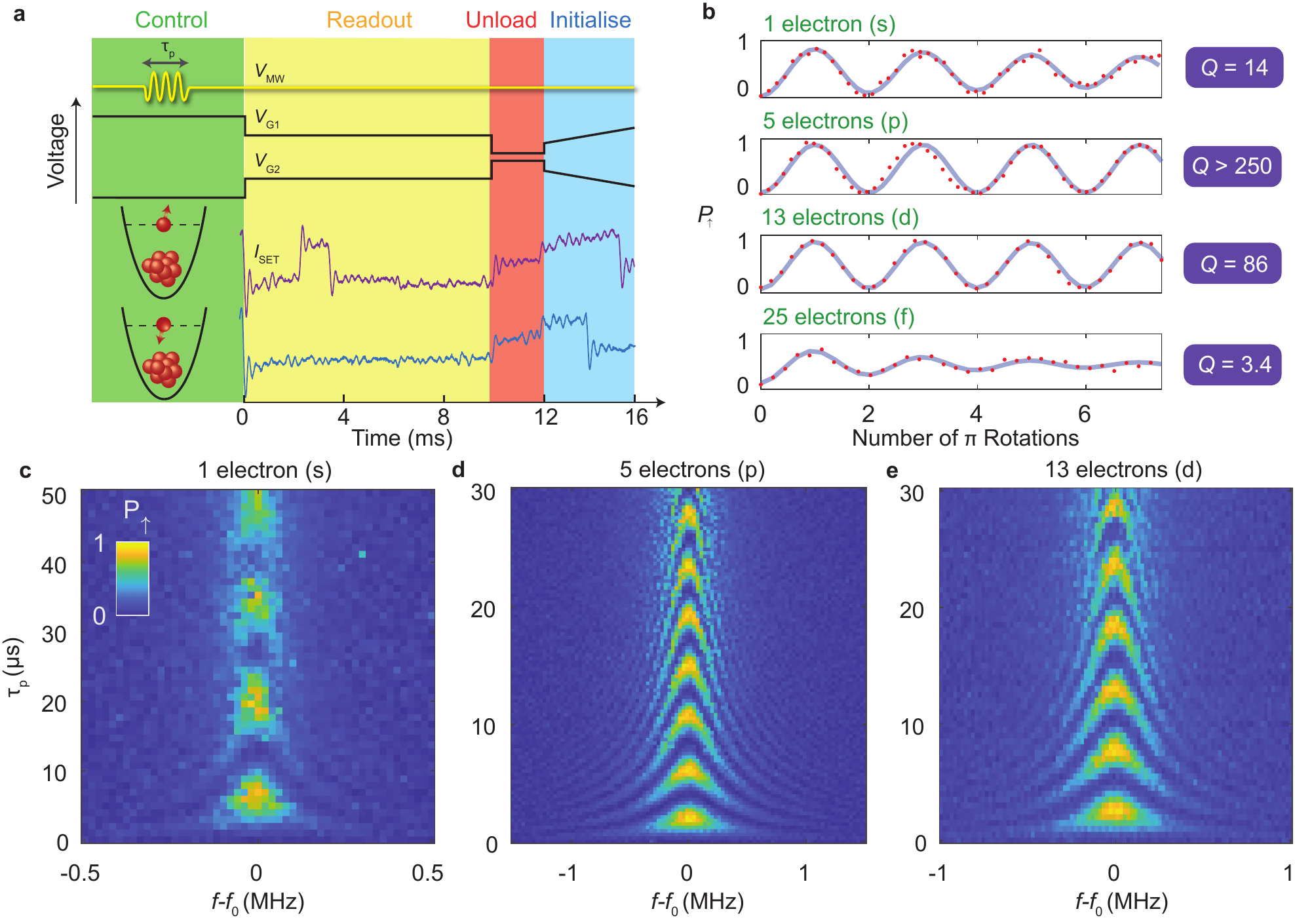}
		\caption{\textbf{Coherent spin control.} 
			\textbf{a}, Gate and microwave pulse sequence for single-qubit control and readout. The lower section shows the change in SET current when a valence electron is in either a spin up or down state.
			\textbf{b}, Rabi oscillation of $N=$1, 5, 13 and 25 electrons. Traces for s, p, d electrons are extracted from \textbf{(c-e)} at $f=f_0$. Horizontal axis is number of $\pi$ rotations ($\frac{\tau_p}{T_\pi}$) of each oscillation. The quality factor $Q=T_2^\textrm{Rabi}/T_\pi$ for each electron occupancy is shown in the right column.
			\textbf{c-e}, Probability of spin up as a function of ESR frequency detuning and duration of microwave pulse for \textbf{(c)} $N=1$, \textbf{(d)} $N=5$ and \textbf{(e)} $N=13$ electrons, performed along the grey dashed line in Fig.~\ref{starkshift}(d-l), which correspond to the highest Q-factor operating points for each electron occupancy. Resonance frequencies $f_0$ for $N=1$, 5 and 13 are 41.829, 41.879 and 41.827 GHz, respectively.
		}\label{SpinControl}
	\end{figure*}
	
	\section*{Operation of single-valence multielectron spin qubits}
	
	We now examine the spins of monovalent dot ocupations as potential qubits, i.e., the first electron of each shell $N=$1, 5, 13 and 25, which we call s-, p-, d- and f-electrons, respectively, in reference to the electronic orbitals~\cite{higginbotham2014coherent}. To demonstrate single-qubit control, we designed this device with the capability to perform electrically-driven spin resonance (EDSR).
	A cobalt micromagnet positioned near the quantum dot induces a magnetic field gradient. An external uniform magnetic field  $B_0=1.4$ T provides a Zeeman splitting between spin states for spin to charge conversion readout~\cite{Elzerman2004}. This field also fully magnetises the micromagnet (cobalt is fully magnetised at $B_0\sim0.4-0.5$T), leading to a field gradient of approximately $1$ T$/\mu$m in the direction transverse to the quantization axis. This provides the means to drive spin flips without the need for an AC magnetic field~\cite{Koppens2006,Veldhorst2014,Pla2012}. Instead, a $\sim$40 GHz sinusoidal voltage is applied directly to the magnet. The antenna-like structure creates an AC electric field at the quantum dot, so that the electron wavefunction oscillates spatially within the slanted magnetic field, which drives Rabi oscillations of the qubit \cite{Pioro-Ladriere2008,Tokura2006,Kawakami2014}.
	
	In order to initialize, control and readout the spins, the pulse sequence depicted in Fig.~\ref{SpinControl}a is performed. The amplitude and duration of the driving AC electric field is used to implement various single-qubit logical gates. The fidelity of these qubit operations under the decoherence introduced by the environment is probed by a randomized benchmarking protocol\cite{Knill2008,Magesan2011}.
	Extended Data Fig.~\ref{extfig-RBM} compares the performance of s-, p- and d-electrons. Single-qubit elementary gate fidelities improve from 98.5\% to 99.7\% and 99.5\% when the electron occupancy increases from 1 to 5 and 13 electrons. Part of the reason for this improvement is the reduction of the quantum dot confinement at higher occupations -- the Coulomb repulsion due to electrons in inner shells leads to a shallower confinement, thus reducing charging and orbital energies (Fig.~\ref{SEMs}d) and ultimately leading to faster Rabi frequencies (see supplementary material).

	A more intuitive way to probe the effects of faster gating times is by measuring the $Q$-factor ($Q=T_2^\textrm{Rabi}/T_\pi$) of Rabi oscillations of 1, 5, 13 and 25 electrons (see Fig.~\ref{SpinControl}b), which shows close to an order of magnitude increase from 1 to 5 and 13 electrons, with a maximum $Q > 250$ for 5 electrons. Moreover, Rabi chevron plots in Fig.~\ref{SpinControl}c-e show a visible improvement in the quality of both $N=$ 5 and 13 electrons compared to $N=$ 1. Further coherence time measurements were also performed, with $T_2^*$ ranging from 5.7 to 18.1 $\mu s$ and $T_2^\textrm{Hahn}$ between 21.6 and 68.5 $\mu s$ (see supplementary material for details)\cite{Kha2015,Zhao2019}.The small variations in coherence are largely compensated by the enhanced Rabi frequency for p and d electrons, which explains the improved qubit performances.
	
	Although Rabi oscillations are visible for $N=25$ in Fig.~\ref{SpinControl}b, we observed its optimal $\pi$-pulse time and $T_2^\textrm{Rabi}$ to be similar to $N=1$. This indicates that higher shell numbers do not necessary benefit qubit operation, as more relaxation hotspots will arise with increased multiplicity of the shell states~\cite{Yang2013}.

	\begin{figure*}
		\centering
		\includegraphics[width=\linewidth]{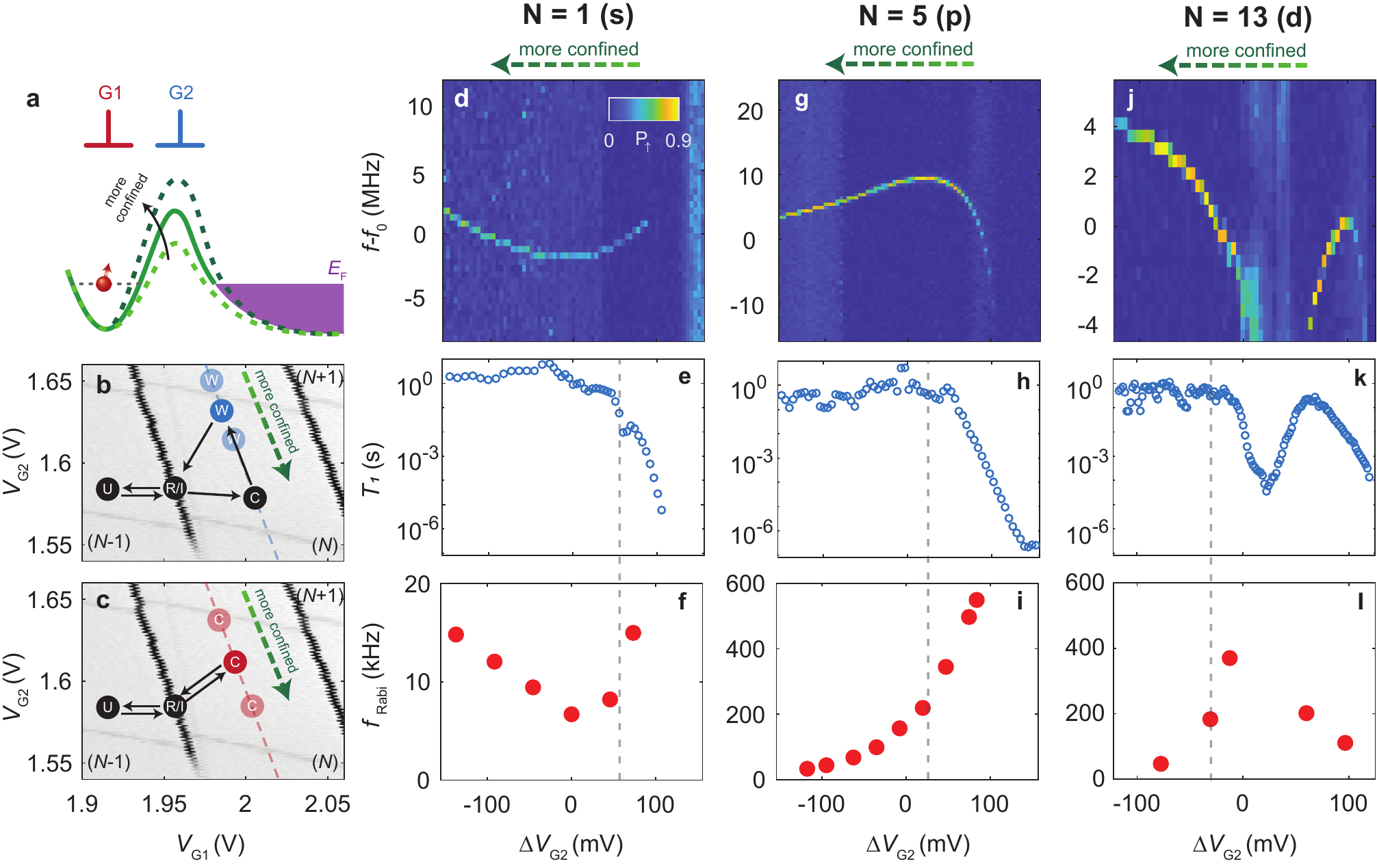}
		\caption{\textbf{Stark shift, tunable Rabi frequency and relaxation time.} 
			\textbf{a}, Schematic representation of the quantum device energy band diagram. G2 voltage varies in order to change the quantum dot size and tunnel rate to the reservoir (purple). Compensating voltage is also applied to G1 to maintain the quantum dot energy level relative to the Fermi level $E_\textrm{F}$.
			\textbf{b,c}, Schematics of the pulse sequences for \textbf{(b)} T$_1$ relaxation, and \textbf{(c)} Rabi control experiment. The qubit control point varies along the dashed line inside the charge stability diagram, parallel to the charge transitions, with electron occupancy either $N=$1, 5 or 13.
			\textbf{d}, Non-linear Stark shift of qubit resonance frequency is observed when the qubit control point changes along dashed line in \textbf{(b)} and \textbf{(c)}, for the $N=1$ electron. At certain voltage levels, the resonance frequency shifts dramatically and eventually qubit readout is unachievable.
			\textbf{e}, Correlation is observed between the magnitude of the differential ESR resonance frequency ($f-f_0$) and qubit relaxation time $T_1$. 
			\textbf{f}, Correlation is also observed between $f-f_0$ and Rabi frequency $f_\textrm{Rabi}$. Maximum Rabi frequency is obtained when the change in ESR frequency $|\frac{\Delta f}{\Delta V_\textrm{G2}}|$ is maximised.
			\textbf{(g-l)}, Stark shift, $T_\textrm{1}$ and Rabi frequencies as plotted in Figs.\textbf{(d-f)}, but for \textbf{(g-i)} $N=5$ and \textbf{(j-l)} 13 electrons. 
			Examples of optimal qubit operation voltages, where a balance exists between fast Rabi oscillation and long spin lifetime, are drawn as grey vertical lines in the figure.
			$f_\textrm{0}=$41.835, 41.870, 41.826 GHz for $N=$1, 5 and 13 electrons, respectively.
		}\label{starkshift}
	\end{figure*}
	
	\section*{Impact of excited states on multielectron qubits} 
	
	Although multielectron quantum dots can be exploited to improve qubit performance, they raise new questions regarding the many-body physics of these dots. One particular concern is that the presence of low-lying excited orbital states may interfere with the spin dynamics. We track the excited states by altering the dot aspect ratio without changing its occupancy\cite{Hwang2017}(see schematic in Fig.~\ref{starkshift}a), by adjusting the G1 and G2 gate voltages as indicated in Fig.~\ref{starkshift}b and c. 
	We first measure the qubit resonance frequencies while varying the dot shape (Fig.~\ref{starkshift}c). This frequency is impacted by variations in g-factor and micromagnet field as the dot is distorted by the external electric field -- we collectively refer to these effects as Stark shift. Linear Stark shift should be observed since the control point of the quantum dot is far detuned from any charge transition. Instead, non-linear Stark shifts are observed for $N=1$ (Fig.~\ref{starkshift}d), $N=5$ (Fig.~\ref{starkshift}g) and $N=13$ electrons (Fig.~\ref{starkshift}j). Although such phenomenon can be partially explained  by change in magnetic field experienced by the quantum dot along the x-direction, a significant drop in resonance frequencies is observed for  $N=5$ (Fig.~\ref{starkshift}g) and 13 electrons (Fig.~\ref{starkshift}j) at $\Delta V_\textrm{G2}>100$mV and 20mV$<\Delta V_\textrm{G2}<60$mV, respectively.
	
	To investigate this further, we measure the spin relaxation time $T_1$ using the pulse sequence in Fig.~\ref{starkshift}b, as shown in Fig.~\ref{starkshift}e,h,k. A clear correlation between the curvature in the Stark shifts and the drop in $T_1$ is similar to previous literature\cite{Yang2013,Srinivasa2013,Borjans2018}. This indicates the presence of an excited orbital or valley state nearby the Zeeman excitation, resulting in a reduction of $T_1$.
	
	Since the virtual excited state (either valley\cite{hao2014electron} or orbital\cite{Amasha2008,rashba2008theory}) plays an essential role in EDSR, the excitation energy directly influences the qubit Rabi frequencies. Performing the pulse sequence in Fig.~\ref{starkshift}c, we observe an enhancement of one order of magnitude for the Rabi frequencies of p and d orbitals (Figs.~\ref{starkshift}f,i,l) correlated to the drop in $T_1$ and curvature of the Stark shift. These are indications that the p and d spins are coupled to excited states of a different nature to those for s electrons. There are no charge transitions (or visible features in the charge stability diagram), indicating that the ground state configuration is left unchanged. Note that some Rabi frequency enhancement is also observed for the $N=$1 electron configuration, but it is an order of magnitude lower than for $N=5$ and 13 electrons.
	
	We may exploit this control over the excitation spectrum to induce fast relaxation on demand for qubit initialization, to operate the qubit where $f_\textrm{Rabi}$ is high, and to store it in a configuration where $T_1$ is long. The power of the EDSR drive only impacts the observed $Q$-factor value up to a factor of 2 (see Supplementary Figure 2b), in contrast to recent observations in depletion mode quantum dot experiments~\cite{Takeda2016} where an order of magnitude difference in $Q$-factors were observed. 
	
	The additional relaxation hotspot around $\Delta V_{G1}=$ 10 mV for the d-shell qubit in Figure~\ref{starkshift}k is most likely due to the increased number of near-degenerate orbitals present, which implies more pathways for qubit relaxation. This near-degeneracy could also be related to why the 14 electron configuration follows Hund's rule to give a $S=1$ ground state~\cite{martins2017negative,Deng2018} (see Fig.~\ref{SEMs}f). We note that these higher total spin states are observed to also be coherently drivable, but a detailed study of these high-spin states exceeds the scope of our present work (see Extended Fig.~\ref{extfig-rabi}c $\&$ d).

	\section*{Conclusions}
	
	The results presented here experimentally demonstrate that robust spin qubits can be implemented in multielectron quantum dots up to at least the third valence shell. Their utility indicates that it is not necessary to operate quantum dot qubits at single-electron occupancy, where disorder can degrade their reliability and performance. Furthermore, the larger size of multielectron wavefunctions combined with EDSR can enable higher control fidelities, and should also enhance exchange coupling between qubits\cite{yang2019silicon}. A multielectron system results in a richer many-body excitation spectrum, which can lead to higher Rabi frequencies for fast qubit gates and enhanced relaxation rates for rapid qubit initialization. Future experiments exploring two-qubit gates using multielectron quantum dots will extend this understanding of electronic valence to interpret bonding between neighboring dots in terms of their distinct orbital states. The controllability of the excitation spectrum should also allow for different regimes of electron pairing, including a	possible singlet-triplet inversion\cite{martins2017negative}, mimicking the physics of paramagnetic bonding \cite{Lange2012}.

	\begin{acknowledgments}
		We acknowledge support from the US Army Research Office (W911NF-17-1-0198), the Australian Research Council (CE170100012), and the NSW Node of the Australian National Fabrication Facility. The views and conclusions contained in this document are those of the authors and should not be interpreted as representing the official policies, either expressed or implied, of the Army Research Office or the U.S. Government. The U.S. Government is authorized to reproduce and distribute reprints for Government purposes notwithstanding any copyright notation herein.
		J. C. and M. P. acknowledge support from the Canada First Research Excellence Fund and in part by the National Science Engineering Research Council of Canada.	K. Y. T. acknowledges support from the Academy of Finland through project Nos. 308161, 314302 and 316551.
		
		The authors declare that they have no competing financial interests.
		
	\end{acknowledgments}
	
	\section*{Author Contributions}
	
	R.C.C.L. and C.H.Y. performed the experiments. J.C.L., R.C.C.L., J.C.C.H., C.H.Y. and M.P.-L. designed the micromagnet, which was then simulated by J.C.L and M.P.-L. J.C.C.H. and F.E.H. fabricated the device with A.S.D's supervision. K.W.C. and K.Y.T. contributed to discussion on nanofabrication process. K.M.I. prepared and supplied the $^{28}$Si epilayer. J.C.C.H., W.H. and T.T. contributed to the preparation of experiments. R.C.C.L., C.H.Y., A.S. and A.S.D. designed the experiments, with J.C.L., M.P.-L., W.H., T.T., A.M. and A.L. contributing to results discussion and interpretation. R.C.C.L., A.S., and A.S.D. wrote the manuscript with input from all co-authors.
	
	\bibliographystyle{naturemag}
	\bibliography{EDSRbib5}
	
\renewcommand\thesuppfig{\textbf{\arabic{suppfig}}}
\renewcommand{\suppfigname}{\textbf{Extended Figure}}
\renewcommand{\tablename}{\textbf{Extended Data Table}}
\renewcommand\thetable{\textbf{\Roman{table}}}


\begin{suppfig*}
	\centering
	\includegraphics[width=\columnwidth]{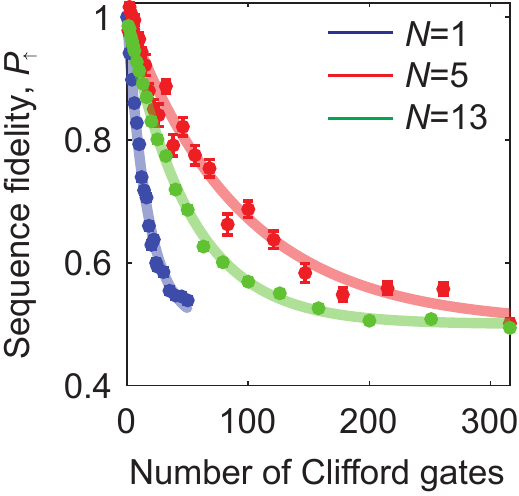}
	\caption{\textbf{Randomised benchmarking.} 
		Single qubit randomised benchmarking at for $N=$ 1, 5 and 13 electrons, with elementary gate fidelities of  98.5 \%,  99.7 \% and 99.5 \%, respectively.
	}\label{extfig-RBM}
\end{suppfig*}

\begin{suppfig*}
	\centering
	\includegraphics[width=\linewidth]{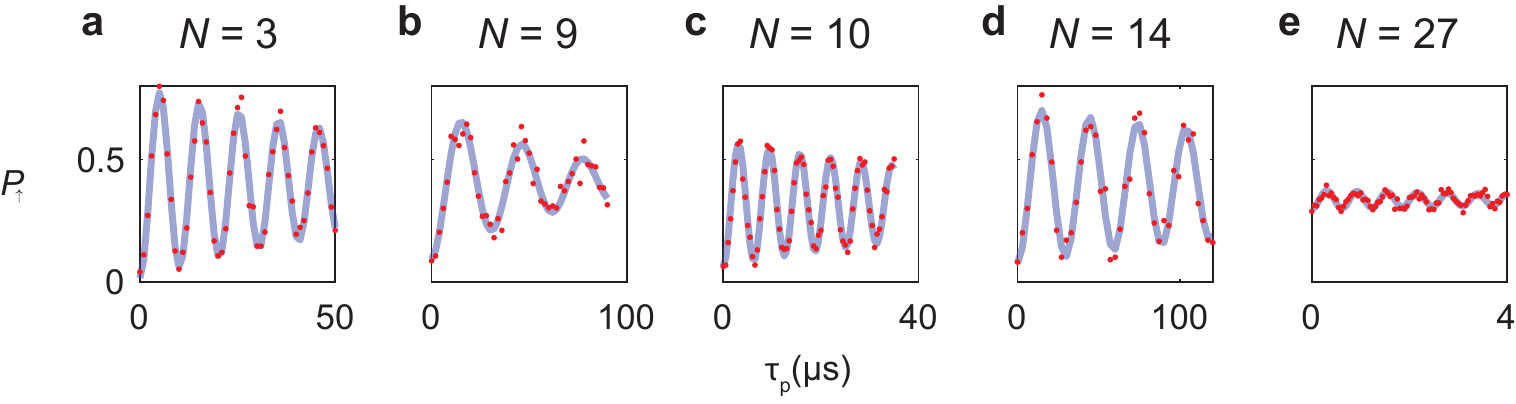}
	\caption{\textbf{Coherent control at various electron occupancies.} 
		Rabi oscillations at different electron numbers $N$ inside the a single quantum dot.
		\textbf{(a)} $N=3$
		\textbf{(b)} $N=9$
		\textbf{(c)} $N=10$
		\textbf{(d)} $N=14$
		\textbf{(e)} $N=27$.
		Note that from Fig.~\ref{SEMs}f, $N=$10 and 14 electrons have total spin states $S=1$, while $N=$27 electrons has $S=\frac{3}{2}$.
	}\label{extfig-rabi}
\end{suppfig*}

\end{document}